\documentclass[aps,prl,floats,twocolumn,floatfix,showpacs]{revtex4}
\usepackage{psfig,epsf,graphicx}
\usepackage{amssymb}
\usepackage{amsmath}
\usepackage{eepic}

\newlength{\piclen}
\begin{document}

\title{Orbital-selective Mott-Hubbard transition in the two-band Hubbard model}

\author{R.\ Arita and  K.\ Held}
\affiliation{Max-Planck-Institut f\"ur
Festk\"orperforschung, 70569 Stuttgart, Germany}

 \date{October 18, 2005}

\begin{abstract}
Recent advances in the field of quantum Monte Carlo simulations
for impurity problems allow --within dynamical mean field theory--
 for a more thorough investigation of the two-band Hubbard model with
narrow/wide band and SU(2)-symmetric Hund's exchange.
The nature of this transition has been controversial,
and we  establish that
an orbital-selective  Mott-Hubbard transition  exists.
Thereby, the wide band still
shows metallic behavior after the narrow band became insulating
-not a pseudogap as for an Ising Hund's exchange.
The coexistence of two solutions with metallic 
wide band  and
 insulating {\em or} metallic narrow
band indicates, in general, first-order transitions. 

\end{abstract}

% 71.27.+a  Strongly correlated electron systems; heavy fermions
% 71.10.Fd  Lattice fermion models (Hubbard model, etc.)
% 71.30.+h  Metal-insulator transitions and other electronic transitions
% 71.20.Eh  Rare earth metals and alloys
% 75.20.Hr  Local moment in compounds and alloys; Kondo effect, valence
%           fluctuations, heavy fermions
% 75.47.Gk  Colossal magnetoresistance

\maketitle

By virtue of dynamical mean field theory (DMFT) \cite{dmft,rmp},
our understanding of the Mott-Hubbard transition \cite{Gebhard} in the one-band Hubbard model
has greatly improved in the last years.
 The bandwidth-controlled
Mott-Hubbard transition is, at least within DMFT \cite{dmftMHT,rmp}, of first-order at low  temperatures  ($T$) and 
becomes a smooth crossover for temperatures above  a critical point,
which terminates the first-order line.
A further complication arises exactly at zero temperature where
two solutions coexist like for low $T$s.
But at $T=0$,  the insulating solution
 is always higher in energy
than the metallic one, i.e.,
the insulating solution is metastable throughout the
whole coexistence region. The  DMFT Mott-Hubbard
transition is of second order at $T\!=\!0$ despite the
coexistence of two solutions.

For making contact with experiments,
orbital realism has to be taken into account, e.g., within the merger
of local density approximation  and DMFT  (LDA+DMFT approach \cite{ldadmft}).
In the case of transition metal oxides, typically 
either the three $t_{2g}$ or the two $e_g$ bands cross the
Fermi energy. At the very least, these orbitals should
be included. For degenerate orbitals,
the situation seems to be clear, at least within DMFT: there is 
a first-order Mott-Hubbard transition \cite{dmftDegOrb}.
Most transition metal oxides are, however, non-cubic.
Hence, the orbital degeneracy is lifted.
Take, for example, the unconventional superconductor
Sr$_2$RuO$_4$ \cite{sr2ruo4}
which has a wide $d_{xy}$ band
and narrow $d_{xz,zy}$ bands \cite{lda} and which 
 becomes a  Mott-Hubbard insulator 
upon substituting Sr by Ca \cite{ca2ruo4}.

For such a situation with wide and narrow 
bands the details of the Mott-Hubbard transitions are so far inconclusive,
even within DMFT and even for a simple two-band Hubbard model with Coulomb
interaction $U$ and Hund's exchange $J$  between the two bands:
Koga {\em et al.} \cite{Koga} employed the so-called exact diagonalization (ED)
method to solve the DMFT equations and obtain 
two Mott-Hubbard transitions:
first the narrow band becomes insulating, then the wide band. This scenario
has been coined orbital-selective Mott-Hubbard transition \cite{OSMHT}.
In contrast, Liebsch \cite{Liebsch} uses  quantum Monte Carlo (QMC) simulations
and the iterated perturbation theory (IPT) to solve the DMFT equations and
finds  a single first-order
Mott-Hubbard transition with similar changes in both bands.
On the insulating side, the wide band has a pseudogap which gradually
amplifies to a real gap with increasing $U$.
In principle, the QMC is more suitable for addressing the 
Mott-Hubbard transition
since ED only gives discrete peaks in the spectra, making
it difficult to unambiguously identify a gap.
However, the QMC simulations are restricted to relatively  high temperatures
and there is a sign-problem \cite{Held} if the Hund's exchange coupling is
taken into account in full, i.e., not only the Ising
but also the spin-flip component.
%Both, Koga {\em et al.} and Liebsch, made 
%the shortcomings of the respective other
%method responsible for the   contradictory results.
Since
the  same limitations  as in \cite{Liebsch} also apply to all previous 
LDA+DMFT(QMC) calculations \cite{ldadmft},
there is  an urgent  need to clarify
whether and how the
details of the  Mott-Hubbard transition are  affected.
Another important aspect is whether
two solutions coexist.  Liebsch finds
two coexisting solutions at a single transition,
while Koga {\em et al.}  \cite{Koga} do not address this question.
If there was a first-order transition 
two consecutive transitions might even be bridged
into a single one.

In this paper, we study this transition
by employing the most recent advances 
in the field of QMC simulations for DMFT.
 The recently introduced  projective QMC (PQMC) method
\cite{Feldbacher}
enables us to address $T\!=\!0$.
Furthermore, the new  Hubbard-Stratonovich decoupling of \cite{Sakai}
allows for 
the calculation with the full SU(2)-symmetric Hund's exchange
at a still-manageable sign-problem, in particular in combination with PQMC.
%With this Paper, we also establish
%the course for future DMFT(QMC)  calculations which
%include the full Hund's exchange coupling.

{\it Model.} Starting point is the two-band Hubbard model
\noindent
%\begin{minipage}{12cm}
\begin{eqnarray}
H&=&-\sum_{m=1}^{2} {t_m} \sum_{\langle i,j \rangle\,\sigma} \hat{c}^\dagger_{im \sigma} \hat{c}^{\phantom{\dagger}}_{jm \sigma}\label{Eq:2bandHubbard}\\&&+\nonumber{U} \sum_{i\, m\sigma}  \hat{n}_{im\uparrow}  \hat{n}_{im\downarrow}
+\!\!\sum\limits_{i; \sigma<\sigma'}\!\!
({U'\!-\!\delta_{\sigma\sigma'}J}) \hat{n}_{i1\sigma\phantom{'}}\!  \hat{n}_{i2\sigma'}\\&&\!
\underbrace{-\frac{J}{2}\! \!{\sum\limits_{{ i \sigma;l\neq m}}}\!\!
\hat{c}^\dagger_{il\sigma }
\hat{c}^{\phantom{\dagger}}_{il\bar{\sigma}}
\hat{c}^\dagger_{im\bar{\sigma}}
\hat{c}^{\phantom{\dagger}}_{im\sigma}
-\frac{J}{2} \! \!{\sum\limits_{{ i \sigma;l\neq m}}}\!\!
\hat{c}^\dagger_{il\sigma }
\hat{c}^\dagger_{il\bar{\sigma}}
\hat{c}^{\phantom{\dagger}}_{im\sigma}
\hat{c}^{\phantom{\dagger}}_{im\bar{\sigma}}}_{\equiv H_2}.\nonumber
\end{eqnarray}
Here,  ${\hat{c}}^{\dagger}_{i m \sigma}$ and
${\hat{c}}^{\phantom{\dagger}}_{i m \sigma}$
are creation and annihilation operators for electrons
on  site $i$ within orbital $m$ and with
spin $\sigma$. The first line describes the kinetic energy
for which we employ the semi-elliptic  non-interacting density of states
 $N^0 (\varepsilon)\!=\!\frac{1}{\pi {W_m^2}/8} \sqrt{(W_m/2)^2-\varepsilon^2}$
(orbital-dependent hopping amplitudes
$t_m$ on a Bethe lattice).
 For the following calculations,
we use different widths for the two bands:
$W_1\!=\!4$ for the wide and  $W_2\!=\!2$ for the narrow band
as in \cite{Liebsch,Koga}. Note that there is no hopping/hybridization between the two bands.
The second line describes the intra- ($U$) and
inter-orbital ($U'$) Coulomb interaction as well
as the Ising-component of the Hund's exchange $J$
($U'\!=\!U\!-\!2J$ by symmetry; we set $J\!=\!U/4$ as in \cite{Liebsch,Koga}).
The third line consists of the spin-flip contribution
to the Hund's exchange (yielding together with the second line 
an SU(2)-symmetric contribution which can also be written as
 $J {\mathbf S}_{i 1} {\mathbf S}_{i 2}$, where ${\mathbf S}_{i m}$ denotes
the spin for orbital $m$ and site $i$).
The last term represents  a pair-hopping term of same strength $J$.

{\it Method.} 
 QMC calculations which take
the spin-flip component of  Hund's
exchange term into account have been a
challenge. Although a straight-forward Hubbard-Stratonovich decoupling,
${\rm exp}(J \Delta\tau c_{1}^\dagger c_{2}c_{3}^\dagger c_{4}) 
\!=\! (1/2) \sum_{s=\pm 1} {\rm exp}[s\sqrt{J\Delta\tau} 
(c_{1}^\dagger c_{2}\!-\!c_{3}^\dagger c_{4})]$,
is possible,  
it has been recognized that it leads to a serious sign 
problem \cite{Held}.
Therefore, it was neglected in 
almost every DMFT(QMC) calculation so far,
including   \cite{Liebsch}.

To overcome this problem, several attempts have been 
made \cite{Motome,Han,Sakai}. Among these,
Sakai, RA, and Aoki proposed a new discrete transformation 
for the spin-flip contribution of the exchange and pair-hopping 
term\cite{Sakai}:
\begin{equation}
  e^{-\Delta \tau H_2} =
  \frac{1}{2}\sum_{r=\pm 1} e^{\lambda r 
  (f_{\uparrow}-f_{\downarrow})}
  e^{a(N_{\uparrow} + N_{\downarrow})+
  bN_{\uparrow}N_{\downarrow}},
\label{Eq:HSdecoupling}
\end{equation}
Here,
%\begin{eqnarray*}
$\lambda \!\equiv\! \frac{1}{2}\log(e^{2J\Delta \tau}\! +\!
 \sqrt{e^{4J\Delta \tau}\!-\!1})$,
  $ a \!\equiv\! -\log(\cosh(\lambda))$, $b \!\equiv\! \log(\cosh(J\Delta \tau))$,
  $f_{\sigma} \!\equiv\! c_{1\sigma}^\dagger c_{2\sigma}\! + \!
  c_{2\sigma}^\dagger c_{1\sigma}$,
  $N_{\sigma} \!\equiv\! n_{1\sigma} \!+\! n_{2\sigma} 
 \! -\!2n_{1\sigma} n_{2\sigma}$.
%\end{eqnarray*}
The advantage of this decoupling is that 
the auxiliary field $r$ is real in contrast 
to that of \cite{Motome}. Hence,
it is expected to yield better statistics in general \cite{Sakai}.

However, even with this decoupling, we note that 
the usual Hirsch-Fye QMC algorithm \cite{HirschFye} does not work
very well for DMFT calculation in the strong coupling regime 
or at low temperatures. For instance,
for  Hamiltonian (\ref{Eq:2bandHubbard})
and $J=U/4$, we found it to be infeasible
to obtain a self-consistent DMFT solution for $U > 2.2$ when 
$\beta (=1/T)>50$ because the Green function $G(\tau)$ has 
a large statistical error at $\tau \sim \beta/2$. 
Therefore, it is difficult to clarify without ambiguity 
whether an orbital selective Mott transition indeed occurs 
in multi-orbital systems at low $T$ by means of 
finite-temperature  Hirsch-Fye QMC calculations;
also see \cite{Koga05}.

Another recent advancement was the development of
 a new projective QMC (PQMC)
algorithm by Feldbacher, KH, and Assaad \cite{Feldbacher}. 
In this algorithm,
ground state expectation values ${\langle\Psi_{0}|\mathcal{O}|\Psi_{0}
\rangle}/{\langle\Psi_{0}|\Psi_{0}\rangle}$ of an
arbitrary operator ${\cal O}$ are calculated as:
\begin{eqnarray}
\langle\mathcal{O}\rangle_{0}
%&=&
%\frac{\langle\Psi_{0}|\mathcal{O}|\Psi_{0}
%\rangle}{\langle\Psi_{0}|\Psi_{0}\rangle}=
%\lim_{\theta\rightarrow\infty}
%\frac{\left\langle \Psi_{T}\right|  
%e^{-\theta H/2}\mathcal{O}e^{-\theta H/2}
%\left|  \Psi_{T}\right\rangle }
%{\left\langle \Psi_{T}\right|  e^{-\theta H}
%\left|  \Psi_{T}\right\rangle} \nonumber \\
&=&\lim_{\theta\rightarrow\infty}
\lim_{\tilde{\beta}\rightarrow\infty}
\frac
{\operatorname*{Tr}
\left[  e^{-\tilde{\beta}H_{T}}e^{-\theta H/2}
\mathcal{O}e^{-\theta H/2}\right]  }
{\operatorname*{Tr}\left[
e^{-\tilde{\beta}H_{T}}e^{-\theta H} \right]  },
\label{Eq:projection}
\end{eqnarray}
where $H_{T}$ is an auxiliary  Hamiltonian
(its ground state  $\left|  \Psi_{T}\right\rangle$
is the trial wave function which is assumed to be non-orthogonal to the ground state $|\Psi_{0}\rangle$ of $H$ \cite{Feldbacher}).

For $H_{T}$, it is convenient to take the one-body part of the
Hamiltonian, because the limit $\tilde{\beta}\rightarrow \infty$ 
can be  taken analytically in this case.
Then, the starting point is the zero-temperature non-interacting
Green function $G_0(\tau,\tau')$  truncated to 
$0\leq\tau,\tau'\leq\theta$ and discretized
as an $L\times L$ matrix ($L=\theta/\Delta\tau$).
From this  $G_0(\tau,\tau')$, 
the  zero-temperature interacting
Green function $G(\tau, \tau')$
is obtained via the same updating equations for the
auxiliary Hubbard-Stratonovich fields
 as for the finite-temperature Hirsch-Fye algorithm. 

While PQMC gives $G(\tau)$  
only for a limited number of (not too large) $\tau$-points, we need $G(i\omega)$ to solve the DMFT 
self-consistency cycle. To this end, the maximum entropy 
method (MEM) is employed to yield the spectral function $A(\omega)$ 
which allows for calculating $G(i\omega_{n}) 
=\int{\rm d}\omega\frac{A(\omega)} {i\omega_{n}-\omega}$ 
at any frequency $i\omega_{n}$. This makes a crucial difference
to finite-temperature calculations.
The large statistical errors occurring
at $\tau\sim \beta/2$ for finite temperatures
  now occur for rather
 large $\tau$'s. But even if there is a large
statistical  error for larger $\tau$'s, the maximum
entropy method can extract sufficient information from
the first  $\tau$ points, discarding the
larger $\tau$'s with excessive statistical error.

One of the main advantages of the PQMC method is that the 
convergence w.r.t.\ $\theta$ is much faster than that w.r.t.\ $\beta$ in 
the Hirsch-Fye algorithm \cite{Feldbacher}.
(Note that the calculation time increases cubically for
$\theta$ and $\beta$.) Hence, we take in the following PQMC calculations
a finite $\theta=20$  ($L=64$), which 
should be sufficiently close to the $T=0$ result:
quantitatively small deviations are  expected for larger $\theta$'s;
qualitatively the behavior should not change anymore
 as in \cite{Feldbacher}.
Similarly as in  \cite{Feldbacher},
the central ${\cal L}=20$ are for measurement
and ${\cal P}=(L-{\cal L})/2=22$ time slices on the right and 
left side of the measuring interval  for projection. 
Typically, we performed 
$7\times 10^6$ to  $3\times 10^7$ QMC sweeps.

\begin{figure}[tb]
\includegraphics[height=8cm]{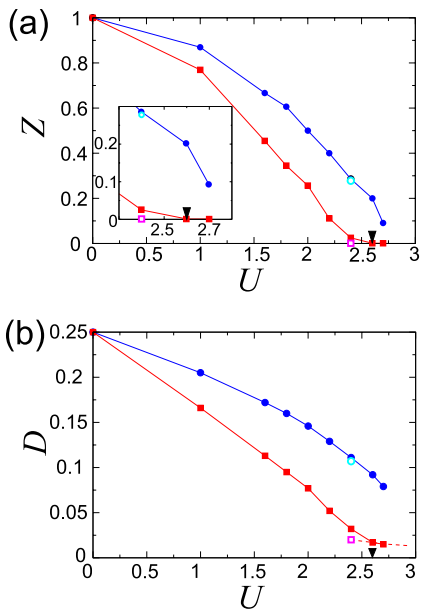}
\caption{
(Color online) (a) Quasiparticle weight $Z$ and (b) double occupancy $D$
as a function of $U$ ($J\!=\!U/4$).
Red (blue) squares (circles) are the data for the narrow (wide) band.
For $U\!=\!2.4$, two solutions are found:
the wide band is metallic for both solutions
whereas the narrow band is metallic (closed symbols)  {\em or} insulating (open symbols).
The solid triangle in (a) and (b) is the $U_c$ estimate from
DMFT(ED) \cite{Koga}; the inset enlarges the behavior around the
transition.
}
\label{D-Z}
\end{figure}

{\it Results.} An indicator for the Mott-Hubbard transition
is the  quasiparticle weight $Z$ which is plotted
in Fig.\ \ref{D-Z}(a).
We clearly see that for the narrow band $Z=0$  for $U \geq 2.6$,
 while  $Z$ is still finite for the wide
band. This means that there is 
a first Mott-Hubbard transition in which
only the narrow band becomes insulating at  $U \approx 2.5$.
This is consistent with the result of 
the DMFT(ED) calculation of \cite{Koga},
in which the critical $U_c$ is estimated to be about 2.6.
In contrast, there is
a single first-order Mott-Hubbard transition at a  smaller value 
$U_c\!\approx\! 2.1$ 
if only the Ising-component of Hund's exchange is taken into account (at $T\!=\!0.03$; between $U_c\!\approx\! 1.8$ and $2.1$
there are two coexisting solutions/hysteresis)\cite{Liebsch}. In our DMFT(PQMC) results,
the wide  band  is still metallic   at $U\!=\!2.7$.
But eventually,
also  the 
wide band has to become insulating at larger Coulomb interactions,
 since in the atomic limit
both bands are insulating. (The calculation for larger Coulomb interactions
unfortunately  became computationally too expensive as even in the PQMC
the statistical error brought about by the spin flip term
of Hund's exchange increases dramatically.) 
Nonetheless, we can conclude from the data available that
there are two different Mott-Hubbard transitions in which 
first the narrow and then the wide band become insulating.
We have an orbital-selective Mott-Hubbard transition.

In Fig.\ \ref{D-Z}(b),  the double occupancy 
$D=\langle n_{\uparrow}n_{\downarrow} \rangle$ for the two different bands
is plotted as a function
of $U$. We see that for the narrow band $D\approx 0.01$ for $U \geq 2.6$.
A similar value of $D \approx 0.01$
was reported \cite{dmftMHT}  for the one-band  Hubbard model 
above the Mott-Hubbard transition, i.e.,  for $U/W \geq 5.9/4$.
This suggests a  Mott-Hubbard
transition very similar to the one-band Hubbard model,
albeit only for the narrow band.

\begin{figure}[tb]
\includegraphics[width=8.6cm]{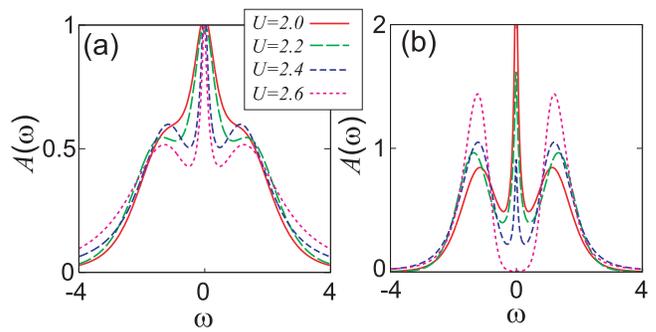}
\caption{  (Color online)
Spectral functions $A(\omega)$ 
for  (a) the wide band and (b) the narrow band. 
For $U=2.6$, the narrow band is insulating while
the wide band is metallic.
}
\label{rho}
\end{figure}

Final evidence for the orbital-selective Mott-Hubbard transition
is obtained from the spectral functions 
shown in
Fig.\ \ref{rho}:
We can unambiguously say that the wide band is still metallic at $U=2.6$,
whereas the narrow band is already insulating with a pronounced gap.
 While the wide band shows a pseudo gap for an
Ising-type of Hund's exchange \cite{Liebsch}, our SU(2) 
symmetric result reveals a metallic peak  in
Fig.\ \ref{rho}.

Let us now study the possibility of first-order 
Mott-Hubbard transitions.
The first question is whether
 at $U\!=\!2.6$ (where we find
a metallic 
wide and insulating narrow band) 
a second solution 
in which both bands are insulating
(co)exists.
Starting the DMFT self-consistency cycle
with an insulating self-energy
for the wide-band, we obtain however
the very same (single) solution 
as in Figs.\ \ref{D-Z} and \ref{rho}.
This demonstrates that the orbital-selective Mott-Hubbard transition
is not merged into a single first-order transition.
There are two distinct Mott-Hubbard transitions.

The second question is,
Are  the  orbital-selective Mott-Hubbard transitions  (generally) 
of first-order? In this case,   
two solutions should coexist for $U\lesssim2.6$: one with a
metallic {\em and} one with an insulating narrow band.
Special care is necessary for insulating 
solutions in the PQMC with a very narrow charge gap.
For such small charge gaps, $\theta$ might not be sufficiently large 
to project --via $e^{-\theta H/2}$-- from the metallic trial wave function
onto the insulating ground state solution.
We then note systematic errors even for intermediate $\tau$'s,
and
substantial noise appears in the charge gap
of the spectrum calculated with the maximum entropy method.
This makes the stabilization of a small-gap insulating
solution delicate. This problem can be mitigated however
by doing the maximum entropy calculation
with a reduced number of $\tau$ points.
Therefore, we used  $\tau$ points up to
$\tau_c \sim 3.2$ and $\sim 1.6$   
for the following results.

\begin{figure}[tb]
\hspace{2cm} wide band \hfill {narrow band\hspace{1.2cm}}

\includegraphics[width=7.8cm]{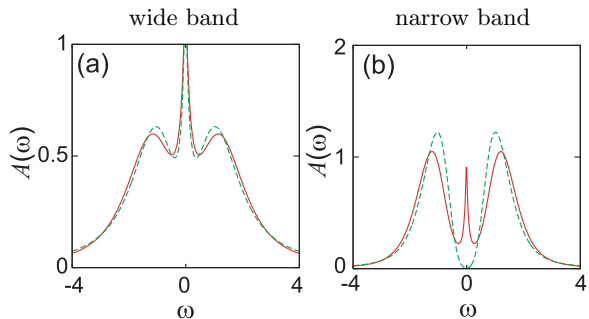}
\caption{ (Color online)
Spectral functions $A(\omega)$ for   (a) the wide and (b) the narrow band at $U=2.4$. 
Two solutions  with insulating/metallic narrow band coexist.
}
\label{U24rho}
\end{figure}

For almost all values of $U$, only a metallic or only 
an insulating solution
is obtained for both $\tau_c \!\sim\! 3.2$ and $\sim\! 1.6$.
However, for $U\!=\!2.4$, we find
 both a solution with a metallic and with
an insulating narrow band (the wide band is metallic in
both solutions with only  minor differences).
In Fig.\ \ref{U24rho}, the spectral function of
these two solutions are shown; the value of $Z$
and $D$ for the insulating solution is plotted in Fig.\ \ref{D-Z}
as open circles and squares.
The DMFT(PQMC) data suggest that two solutions with
metallic and insulating narrow band
coexists for $U \!\sim\! 2.4$, so that the Mott-Hubbard transition 
in which the narrow band becomes insulating (and in which 
the wide band stays metallic)
 is in general of first-order. Possibly, the insulating
solution is metastable at $T\!=\!0$.

{\em Discussion.} For understanding the DMFT results it
is instructive to remind ourselves of what is known for the two-orbital
Anderson impurity model (AIM).
If $J>T_K$ (the AIM Kondo temperature)
the impurity spins of the two orbitals form a steadfast  spin-1 (triplet).
For such an AIM and inequivalent orbitals
 it is known that this spin-1 is screened in two stages:
first only by one orbital to a   spin-$\frac12$ at $T_K^1$, and then by the 
second orbital to a spin-0 at $T_K^2$ \cite{AIM}.
Within DMFT we now have to solve AIMs self-consistently:
The AIM's $T_K$'s of one DMFT iteration
(crudely $T_K \!\approx\! Z W$) sets the hybridization strength 
for the next DMFT iteration.
Hence, we can interpret our DMFT results as following:
Given the two inequivalent Kondo scales of the AIM,
there is a $U$-interval where only the
hybridization strength (and $T_K$) of  the narrow orbital
is renormalized to zero by the DMFT iterations. Only the narrow band is insulating.

If only the Ising-component of Hund's exchange is taken into account,
the behavior of the AIM is completely different. Instead
of a triplet, the impurity spins allign to  
$S_Z\!=\!\pm 1$ (no $S_Z\!=\!0$ component). 
For  $J\!>\!T_K$ ($J\!\approx\! 0.5$ and $T_K\!\approx\! ZW\!\approx\! 0.4$ at the Mott-Hubbard transition of \cite{Liebsch}), there is no spin Kondo
effect any more since it requires a spin-flip of the conduction electrons
and, hence, a {\em change} of $S_Z$ by $\pm 1$. As soon as one orbital becomes
insulating, there is also no orbital Kondo effect anymore: the whole
system is unscreened, i.e.,
insulating. It is certainly interesting to study whether
this kind of
physics is relevant for magnetically anisotropic materials.

{\em Conclusion.} Taking the full SU(2)-symmetry of Hund's exchange into account in the PQMC calculation, we conclude that there
are two consecutive Mott-Hubbard transitions,
whereby -at least around the first transition- 
two solutions coexist.
%We also paved the way for future multi-orbital DMFT
%calculations with the full Hund's exchange \cite{note}.
By clarifying the theoretical side, 
we hope to stimulate further experiments on the
 orbital-selective Mott-Hubbard transition,
e.g., in Sr$_2$RuO$_4$ where results were so-far negative in this respect
\cite{Wang}.

We acknowledge very fruitful discussions with M.\ Feldbacher, S.\ Sakai,
and A.\ Toschi
as well as support by the Alexander von Humboldt foundation (RA) and
the Emmy Noether program of the Deutsche Forschungsgemeinschaft (KH).

During the completion of our work, we learned
about several related studies \cite{Koga05,recent}.

\end{document}